\documentclass[onecolumn,preprintnumbers,amsmath,amssymb,superscriptaddress]{revtex4}
\usepackage{graphicx,subfigure,epsfig,epstopdf}
\usepackage{graphicx}
\usepackage{dcolumn}
\usepackage{bm}
\usepackage{amsmath,amssymb,upgreek}
\input{epsf}
\input{epsfx}

\makeatletter

\newcommand{\Rmnum}[1]{\expandafter\@slowromancap\romannumeral #1@}
\makeatother

\begin{document}
\title{High-Contrast Electro-Optic Modulation of a Photonic Crystal Nanocavity by Electrical Gating of Graphene}

\date{October 10, 2012}

\author{Xuetao Gan}
\affiliation{Department of Electrical Engineering, Columbia University, New York, NY 10027}
\author{Ren-Jye Shiue}
\affiliation{Department of Electrical Engineering, Columbia University, New York, NY 10027, USA}
\author{Yuanda Gao}
\affiliation{Department of Mechanical Engineering, Columbia University, New York, NY 10027, USA}
\author{Kin Fai Mak}
\affiliation{Department of Physics, Columbia University, New York, NY 10027, USA}
\author{Xinwen Yao}
\affiliation{Department of Electrical Engineering, Columbia University, New York, NY 10027, USA}
\author{Luozhou Li}
\affiliation{Department of Electrical Engineering, Columbia University, New York, NY 10027, USA}
\author{Attila Szep} 
\affiliation{Air Force Research Laboratory, Sensors Directorate, WPAFB, Dayton, OH 45433, USA}
\author{Dennis Walker Jr.} 
\affiliation{Air Force Research Laboratory, Sensors Directorate, WPAFB, Dayton, OH 45433, USA}
\author{James Hone}
\affiliation{Department of Mechanical Engineering, Columbia University, New York, NY 10027, USA}
\author{Tony F. Heinz}
\affiliation{Department of Electrical Engineering, Columbia University, New York, NY 10027, USA}
\affiliation{Department of Physics, Columbia University, New York, NY 10027, USA}
\author{Dirk Englund}
\affiliation{Department of Electrical Engineering, Columbia University, New York, NY 10027}
\affiliation{Department of Applied Physics and Applied Mathematics, Columbia University, New York, NY 10027}

\begin{abstract}
We demonstrate a high-contrast electro-optic modulation of a photonic crystal nanocavity integrated with an electrically gated monolayer graphene. A high quality ($Q$) factor air-slot nanocavity design is employed for high overlap between the optical field and graphene sheet. Tuning of graphene's Fermi level up to 0.8~eV  enables efficient control of its complex dielectric constant, which allows modulation of the cavity reflection in excess of 10~dB for a swing voltage of only 1.5~V. We also observe a controllable resonance wavelength shift close to 2~nm around a wavelength of 1570~nm and a $Q$ factor modulation in excess of  three. These observations allow cavity-enhanced measurements of the graphene complex dielectric constant under different chemical potentials, in agreement with a theoretical model of the graphene dielectric constant under gating. This graphene-based nanocavity modulation demonstrates the feasibility of high-contrast, low-power frequency-selective electro-optic nanocavity modulators in graphene-integrated silicon photonic chips.  
\end{abstract}
\maketitle

Graphene has intriguing optical properties and enables a range of promising optoelectronic devices~\cite{Bonaccorso2010, Konstantatos2012, Mueller2010, Xia2009,Lemme2011, Liu2011d,  Bao2009,Sun2010, Hendry2010}. To enhance the inherently weak light-matter interaction in this single atomic layer material, grahene has been coupled to optical waveguides and cavities ~\cite{Liu2011d, Liu2012d, Furchi2012, Engel2012,Gu2012}. In the limit of wavelength-scale confinement, we recently demonstrated a dramatic enhancement of the  light-matter interaction for graphene coupled to a planar photonic crystal (PPC) nanocavity, which reduced the cavity reflection by more than 20~dB~\cite{Gan2012}.  Here, we employ this system to demonstrate a high contrast electro-optical modulation of the cavity reflection, in excess of 10~dB. The modulation is achieved by electrical gating of the graphene monolayer using an electrolyte, which, while slow, shows the fundamental capability of graphene-based modulation of nanocavities.  Furthermore, we employ the strong coupling between the cavity modes and graphene for precision spectroscopy of graphene under gating conditions. We measure a complex dielectric constant that agrees with a theoretical model of the optical conductivity in graphene and supports the notion of residual absorption even when the Fermi level is tuned far above the transparency condition for infrared photons at energy $\nu$, namely $E_F>h\nu/2$.

\begin{figure}
\includegraphics[width=6in]{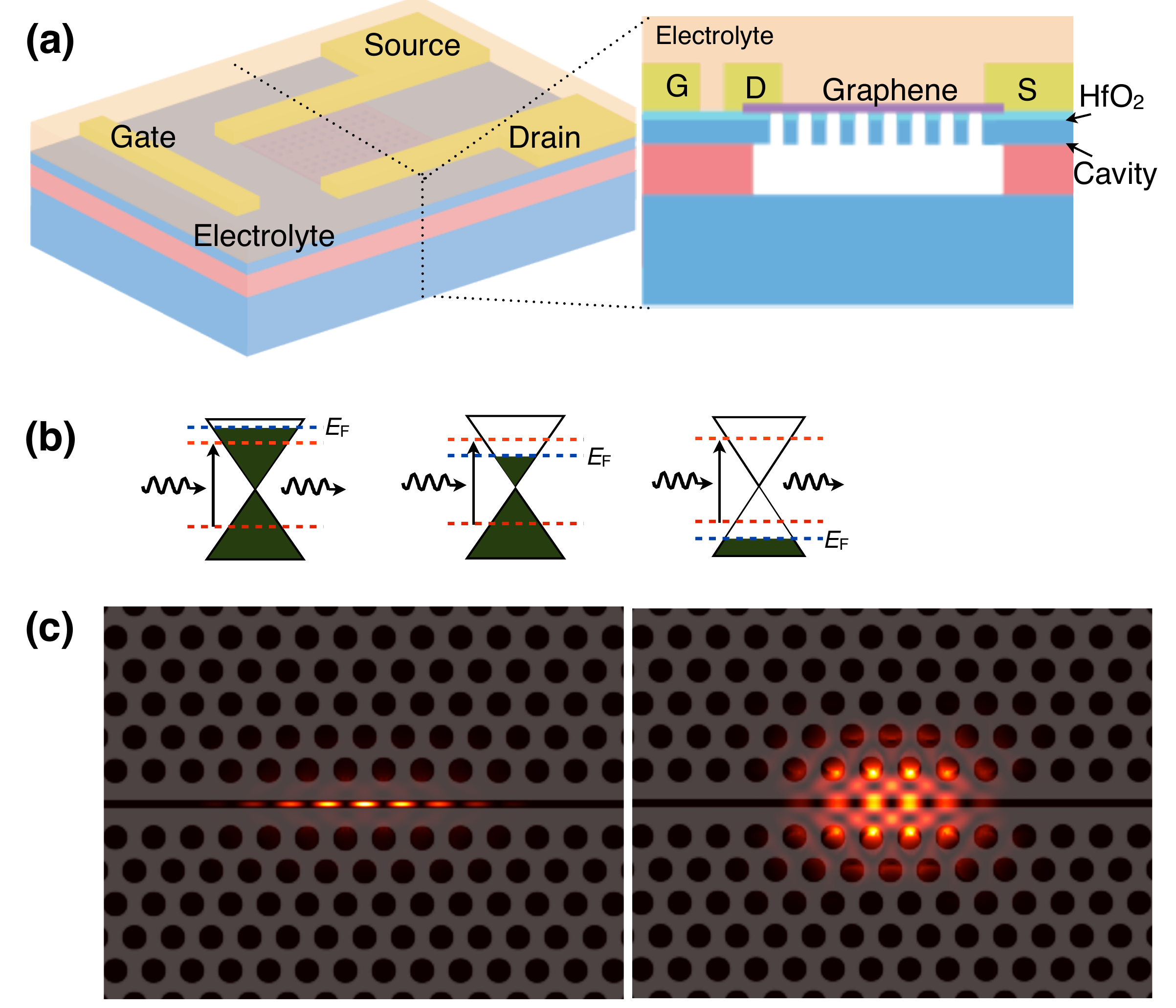}
  \caption{(a) Schematic of the electrically controlled graphene-PPC nanocavity. The cavity reflection can be modulated by electrostatic tuning graphene Fermi level with the top electrolyte. (b) The band structure of graphene with different doping level, where the first and third ones corrpond to the transparent graphene with the inhibited interband transition. (c)  Simulated energy distribution of two resonant modes of the air-slot cavity. The optcial field is confined in the air-gap, enabling a strong coupling with graphene.}
  \label{fgr:example:2}
\end{figure}
\vspace{12pt}

 As shown in Fig.1(a), the experimental device consists of an air-suspended PPC cavity that is coupled to a  graphene field effect transistor (FET) gated by a solid electrolyte ~\cite{Mak2009, Lu2004}. The optical transmission of graphene for an incident photon with frequency $\nu$ is modulated by electrostatic tuning the graphene's Fermi energy ($E_F$). As shown in Fig. 1(c), when $E_F$ is tuned away from the Dirac point by more than half of the photon energy $\nu/2$, the interband transitions are inhibited, reducing graphene absorption~\cite{Wang2008, Li2008b}. 

To improve the overlap between graphene and cavity resonant modes, we employ an an air-slot PPC nanocavity ~\cite{2010.APL.CWW.air_slot} with strongly confined modes in the air-gap, as shown in Fig. 1(b). This design improves the graphene-optical mode coupling rate by approximately two-fold compared to inear three-hole defect cavities used previously~\cite{Gan2012}, where light is confined in the high index material~\cite{Noda2003Nature} and therefore experiences less overlap with the graphene layer.  The air-slot PPC nanocavities are fabricated on a silicon-on-insulator wafer with a  220~nm- thick silicon membrane, using a combination of electron beam lithography and dry/wet etching steps. We employ mechanically exfoliated graphene monolayers, which are transferred onto  PPC nanocavities using a precision alignment  technique~ \cite{Dean2010}. The drain, source, and gate electrodes of the graphene FET are fabricated using electron beam lithography and titanium-gold electron beam evaporation. In previous experiments, we found that these contacts can gate the intrinsic or lightly doped silicon membrane directly and influence the cavity spectroscopy under doping. To avoid this, the devices described in this study include a conformal 10~nm hafnium oxide (HfO$_2$) layer grown on the PPC using atomic layer deposition before the metal contacts are fabricated (see Figure 1(a)).

Figure 2(a) displays an optical image of one of the completed graphene-PPC nanocavity devices. The dashed red line indicates the boundary of the monolayer graphene, which is furthermore confirmed using micro-Raman spectroscopy. The gate electrode is about 15~$\upmu$m from the graphene flake to ensure effective doping through the electrolyte. Figure 2(b) shows a scanning electron microscope (SEM) image of the slot cavity with lattice spacing of $a=450$~nm and lattice hole-radii of $r=150$~nm.  After graphene is transferred and contacted, we spin-coat electrolyte (PEO plus LiClO$_4$) on the full wafer, which provides a high electric field and carrier density in graphene \cite{Mak2009, Lu2004}, while also capping the device. 

\begin{figure}
\includegraphics[width=6in]{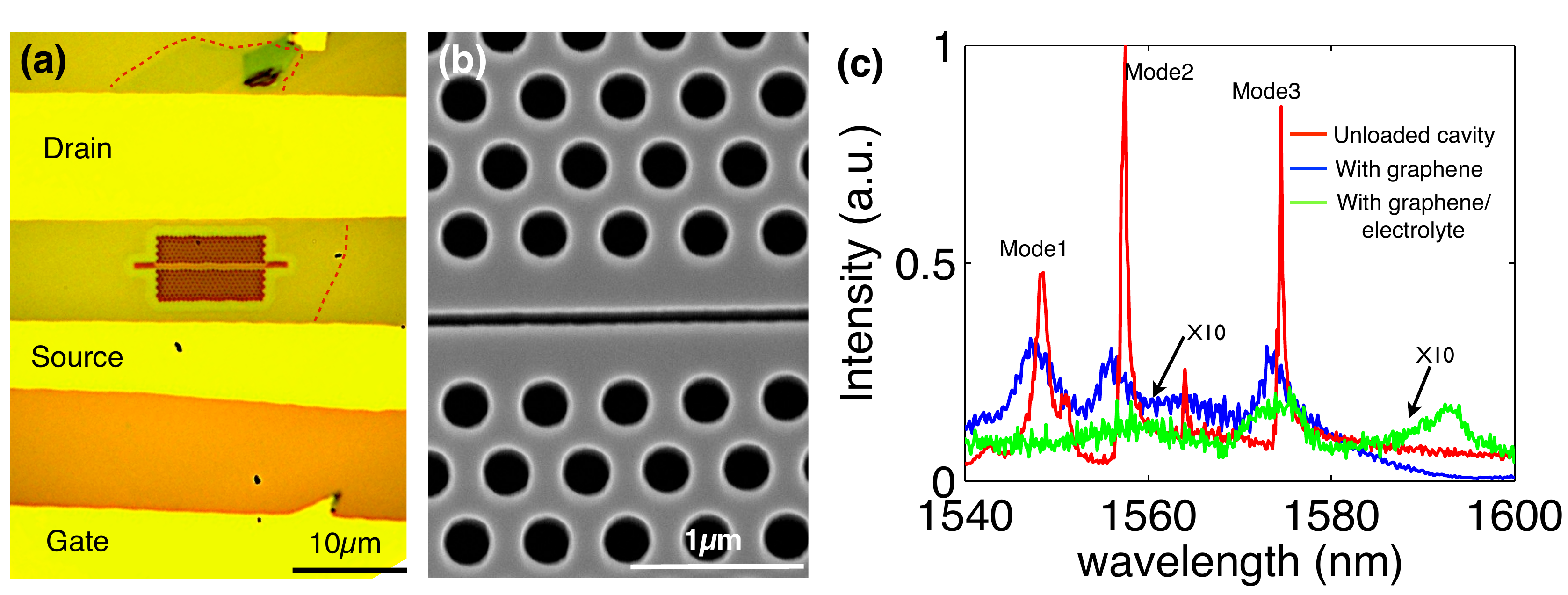}
  \caption{(a) Optical image of one of the electrically controlled graphene-PPC nanocavity devices. The monolayer graphene covers on an air-slot cavity, which is indicated by the red dashed line. Source and drain electrodes are near the cavity, while the gate electrode is about 15~$\upmu$m removed. (b) SEM image of the air-slot cavity before graphene deposition; AFM and SEM studies of PPC cavities after graphene deposition are shown in previous work~\cite{Gan2012} . (c) Reflection spectra of the intrinsic PPC cavity, after transfer of graphene, and after deposition of the electrolyte gate. }
  \label{fgr:example:2}
\end{figure}
\vspace{12pt}
We characterize the graphene-PPC nanocavity  using a cross-polarization confocal microscope with a broad-band (super-continuum laser) excitation source~\cite{Gan2012, 2007.Nature1_etal}. The reflection is analyzed using a spectrometer with resolution of 0.05~nm. The slot cavity has three dominant resonant modes ~\cite{2010.APL.CWW.air_slot} at wavelengths of $\lambda_1=1548.4$~nm (Mode1), $\lambda_2=1557.4$~nm (Mode2), and $\lambda_3=1574.5$~nm (Mode3). As shown in Figure 2(c), the intrinsic cavity resonances first blue-shift upon HfO$_2$ and graphene transfer, as expected for a decrease in the refractive index of the thin HfO$_2$ layer during the annealing step of the graphene transfer process (see Supporting Information). The electrolyte has a real dielectric constant of $\sim 2.1$ that subsequently red-shifts the cavity~\cite{J.D.JoannapolousS.G.JohnsonJ.N.Winn2008}, as shown in the green curve in Fig. 2(c).  Modes 1 and 2 become indistinguishable after graphene and electrolyte deposition because they experience different red-shifts due to different overlap with the electrolyte (see Supporting Information).

Fitting the cavity resonances to Lorentzian curves, we estimate $Q$ factors of $Q_{i}=858, 2350, 3420$ for modes $i=1, 2, 3$. After the single-layer graphene transfer, the $Q$ factors decrease to 350, 640, and 440, respectively. Employing perturbation theory to the graphene-PPC system~\cite{Gan2012}, we obtain that the energy decay rates due to graphene absorption for these three resonant modes are $\kappa_ {cg}=(1.6, 1.1, 2.0)\times 10^{-3}\omega_i$, where $\omega_i=2\pi c/\lambda_i$ are the resonant frequencies of the cavity modes. These loss rates indicate that the coupling between the graphene and mode3 is strongest, which is expected from the  electric field distributions of the resonant modes: modes 1 and 3 have a strong optical fields --- and therefore a large overlap with graphene ---in the air-gap, while the overlap with the air slot is far weaker for mode 2. The single-layer graphene causes a strong (nearly 18~dB) reduction of the reflected intensity in the three resonant peaks. After electrolyte deposition, the $Q$ factors of the resonances drop slightly to 300 and 420 for modes 1 and 3, respectively. 

\begin{figure}
\includegraphics[width=6in]{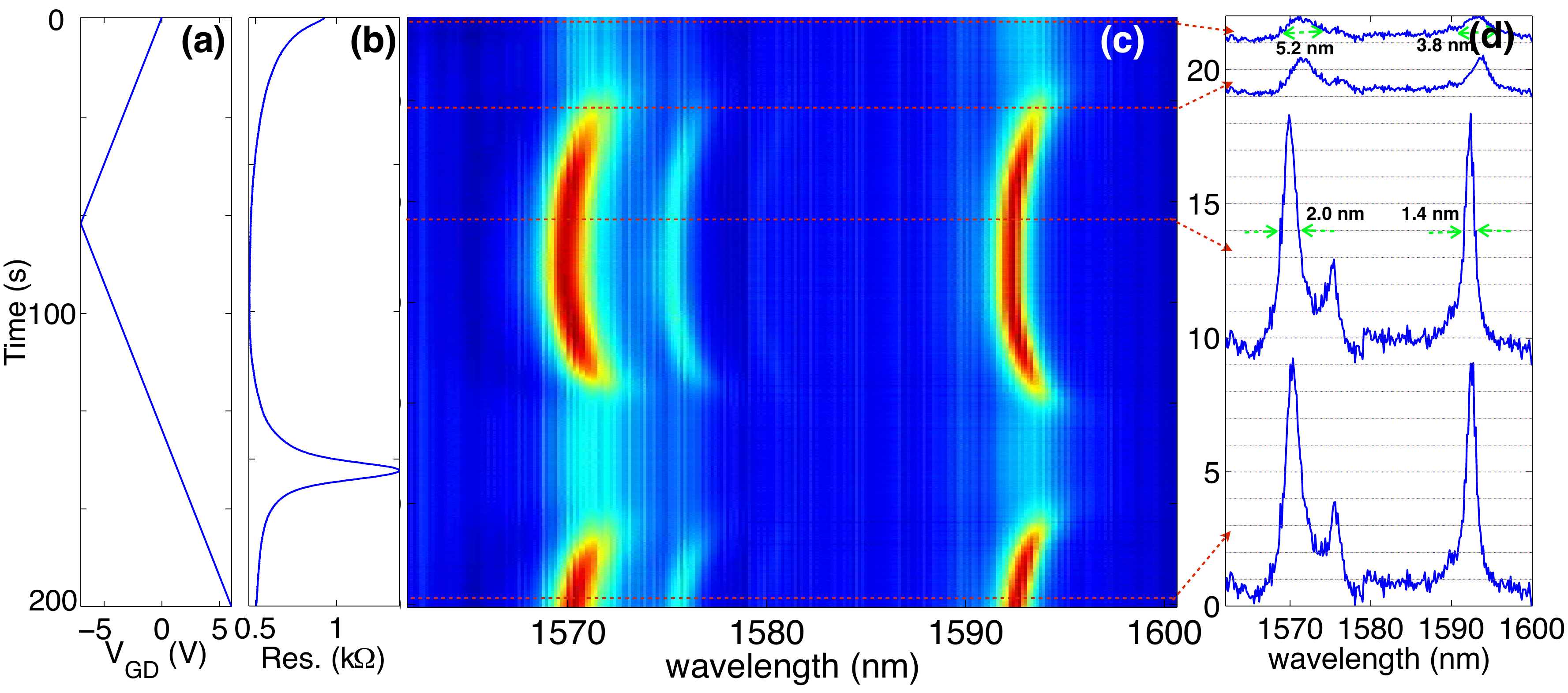}
  \caption{Electrical and optical response of the electrically controlled graphene-PPC nanocavity. (a) The gate voltage $V_g$ modulated in a saw-tooth pattern at a rate of 0.1~V/s between -7~V and 6~V. (b) DC Resistance across the graphene layer measured from the source to drain electrodes, showing a charge neutral point at $V_{CN}=1.4$~V. (c) Reflection spectra of the cavity as $V_g$ is modulated. The resonant peaks present clear wavelength shift, $Q$ factor and intensity variations during the modulation. (d) Spectra of the cavity reflection for $V_g$=0, -2, -7, and 6~V, which are normalized by the reflection peak at $V_g$=0. Compared to the reflection in the  cavity at zero-bias, the cavity presents $\sim$8.5 times higher peaks when the monolayer graphene is doped highly. }
  \label{fgr:example:2}
\end{figure}
\vspace{12pt}

To study the electrical control of the graphene-PPC nanocavity, we measure the cavity reflectivity  as a function of  the gate voltage $V_g$ across the gate and drain electrodes. To aviod degradation of the electrolyte, the sample is measured in a chamber with vacuum of $10^{-4}$~mbar at room temperature. The electrical signal through the drain and source is monitored simultaneously to record the doping level of graphene. Figure 3 shows the measurements of the electrical and optical signals as the gate voltage $V_g$  is linearly modulated.  Here we scan $V_g$ at a speed of 0.1~V/s between -7~V~---~6~V, as shown in Fig. 3(a). The measured resistance on the graphene FET in Figure 3(b) indicates a gate voltage for the charge neutral point at $V_{CN}=1.4$~V. In these sweeps, the cavity reflection spectra are acquired continuously at frames of  33~ms each, as shown in Fig. 3(c). When $V_g$ is close to  zero (-1~V< $V_g$<0~V), the doping level on graphene is low. The cavity reflection is constant as that in the zero-bias condition, which is shown in the first panel of Fig. 3(d) with two resonant peaks at the wavelengths of 1571.7~nm and 1593~nm for modes 1 and 3, respectively. Tuning the graphene Fermi level by decreasing $V_g$ to -1~V,  we observe that the two resonant peaks become narrower while red-shifting slightly, as shown in the second panel of Fig. 3(d). As the cavity loss is reduced, the cavity reflection intensity increases. Decreasing $V_g$ further,  the resonant peaks grow and narrow sharply over a voltage range of $\sim1.5$~V, while the center wavelengths blue-shift. Finally, the cavity linewidths and intensities saturate as $V_g$ drops below -2.5~V. However, the cavities continue to blue-shift over down to -7~V. The third panel of Fig. 3(d) shows the reflection spectrum at $V_g$=-7~V.  Because the peaks are now very narrow compared to the zero-bias case,  the modes 1 and 2 again become distinguishable. We observe the reciprocal tuning phenomena when we increase $V_g$ back from -7~V to 0~V. For $V_g>0$,  the data show a wide flat spectrum ($V_g$) around the charge neutral point of graphene due to the low doping level. The fourth panel of Fig. 3(d) plots the spectrum at $V_g$=6~V, showing the strong modulation of cavity reflectivity with the high electron side doping on graphene. The resonant peaks are increased to 8.5 when graphene is highly doped compared to zero-bias. Combining with the resonant wavelength shift, the modulation depth at is higher than 10~dB at the wavelenght of 1592.6~nm, which is the resonant wavelength  of mode3 at $V_g$=-7~V.

Fitting modes 1 and 3 in Fig. 3(c) to Lorentizan, we obtain the $Q$ factors and resonant wavelengths as a function of $V_g$  from -6~V to 6~V. These are plotted in Figs. 4(a) and 4(b), respectively.  For both cavity resonances, the changes of the $Q$ factors and resonant wavelengths are symmetric with respect to $V_{CN}$. Due to the higher photon energy of mode1, its $Q$ factor begins to increase after that of mode3. Both resonant modes show the saturation of the $Q$ factor with the change from 300 (420) to 780 (1150) for mode1 (3). Figure 4(b) depicts the initial resonant wavelength red-shift, followed by the mode blue-shift, with a maximum change of 2~nm and 1.3~nm for modes 1 and 3, respectively. 

The observed changes of $Q$ factors and resonant wavelength shifts are due to the change of the complex dielectric constant of graphene~\cite{Gan2012}. We can calculate these changes from a knowledge of the overlap of the cavity modes with the graphene layer and using perturbation theory~\cite{Gan2012}. The results calculated from mode1 are shown in Fig. 4(c). The imaginary part of the graphene dielectric constant has a value of 8.2 at $V_{CN}$. It decreases to the minmum at $V_g-V_{CN}$=3~V, corresponding to $E_F\sim0.4$~eV,  without any change even under higher $V_g$. The real part of the graphene dielectric constant increases monotonically first from 2.1 at the charge neutral point to the maximum value of 3.6 at $V_g-V_{CN}$=1.4~V. Then it decreases to -5.9 continusly under the highest doping level.  Figure 4(d) displays the calculated Fermi level on the graphene sheet as a funtion of $V_g$. The obtained complex dielectric constant of graphene from the experiment measurements matches with theeoretical predictions ~\cite{Lu2012b}.

\begin{figure}
\includegraphics[width=6in]{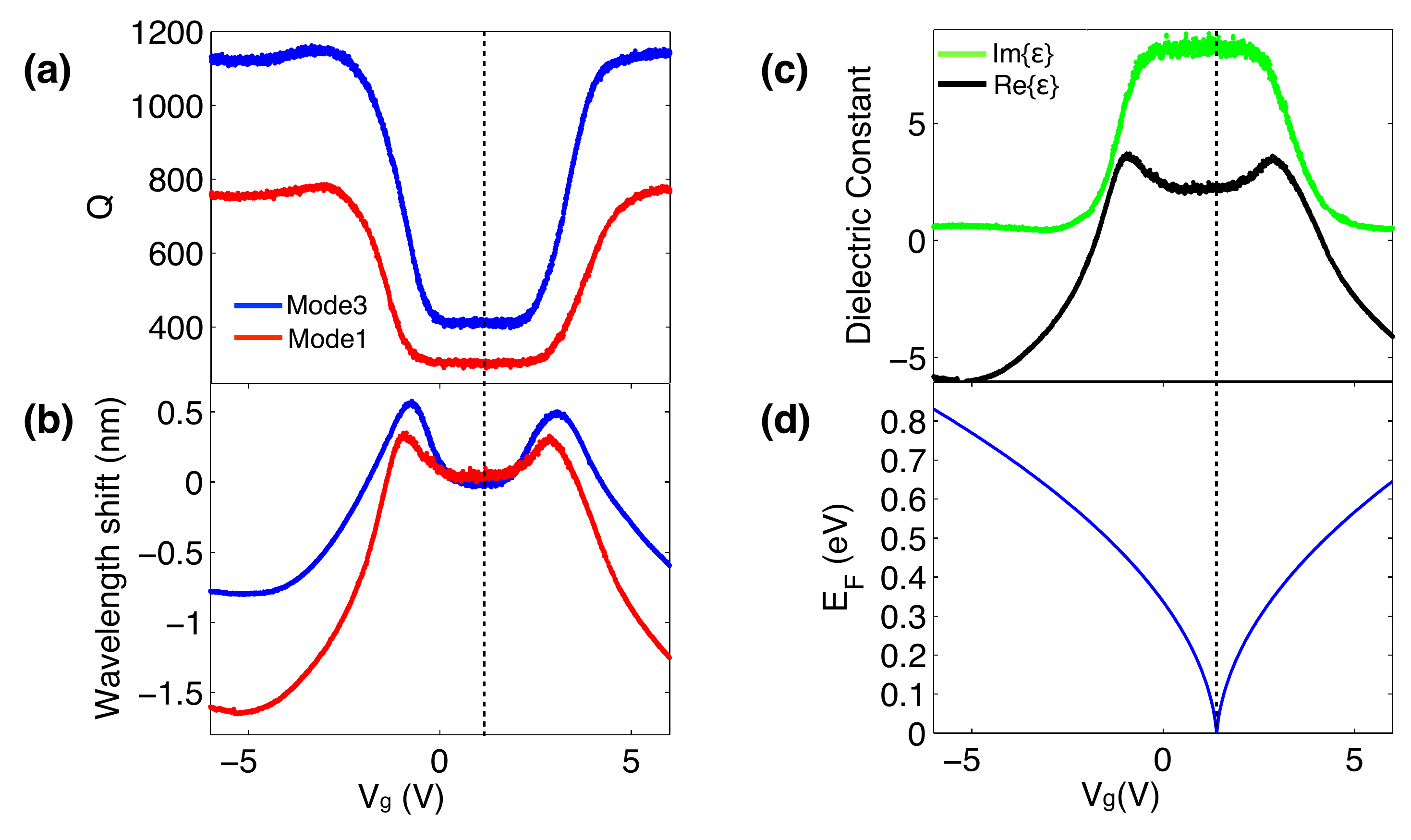}
  \caption{ (a) Change of the Q factors and (b) wavelength shift for Mode1 and Mode3 as a function of $V_g$. (c) Complex dielectric constant of graphene as a function of $V_g$, which is calculated from the resonance modulation in (a) and (b).  (d) Fermi level of the graphene monolayer tuned by $V_g$. }
  \label{fgr:example:2}
\end{figure}

\vspace{12pt}

In conclusion, we have demonstrated that by coupling electrically gated graphene to a PPC nanocavity, it is possible to realize strong optical modulation . This finding shows great promise for future electro-optic chip-integrated, small-footprint modulators that employ doping graphene as the active medium. In future studies, graphene can be back-gated using highly doped silicon PPC cavity or dual graphene layers to increase the modulation into the GHz regimes~\cite{Liu2011d, Liu2012d}. Because of the high mobility and small capacitance of such PPC-graphene devices, we also anticipate low power consumption. The cavity enhancement also enabled the precise measurement of the graphene conductivity in a deep sub-wavelength region of graphene-mode overlap. The cavity therefore enable precision spectroscopy in small graphene volumes.

Acknowledgement:  Financial support was provided by the Air Force Office of Scientific Research PECASE, supervised by Dr. Gernot Pomrenke and by the National Science Foundation through grant DMR-1106225. Fabrication of the PPC was carried out at the Center for Functional Nanomaterials, Brookhaven National Laboratory, which is supported by the U.S. Department of Energy, Office of Basic Energy Sciences, under Contract No. DE-AC02-98CH10886. Device assembly, including graphene transfer, and characterization was supported by the Center for Re-Defining Photovoltaic Efficiency Through Molecule Scale Control, an Energy Frontier Research Center funded by the U.S. Department of Energy, Office of Science, Office of Basic Energy Sciences under Award Number DE-SC0001085.

\bibliography{/Users/xuetao/Documents/library.bib}

\end{document}